\documentclass[12pt,onecolumn]{aastex6}

\newcommand{\filter}[1]{\mbox{\it #1\/}}              % filter in italics
\newcommand{\hcoa}{$\mathrm{HCO^{+}}(1-0)$}
\newcommand{\hcob}{$\mathrm{H^{13}CO^{+}}(1-0)$}
\newcommand{\HII}{HII}
\usepackage{graphicx}

\usepackage{txfonts}
\usepackage{natbib}

\begin{document}

   \title{The stellar content of the infalling molecular clump G286.21+0.17}

   \author{M. Andersen}
          \affil{Gemini Observatory, Casilla 603, La Serena, Chile\email{manderse@gemini.edu}}
        
        \author {P. J, Barnes} \affil{    Department of Astronomy, University of Florida, Gainesville, FL 32611, USA} \affil{School of Science and Technology, University of New England, Armidale NSW 2351, Australia}
         
         \author{J. C. Tan} \affil{ Departments  of Physics \&\ Astronomy, University of Florida, Gainesville, FL 32611, USA}
           
          \author{J. Kainulainen} \affil{ Max-Planck-Institut f\"ur Astronomie, K\"onigstuhl 17, 69117 Heidelberg, Germany}
          
            \author{G. de Marchi} \affil{ Research \& Scientific Support Department, ESA ESTEC, Keplerlaan 1, 2200 AG Noordwijk, The Netherlands}

%   \date{Received September 15, 1996; accepted March 16, 1997}

  \begin{abstract}
     {The early evolution during massive  star cluster formation is still uncertain.}
     {Observing  embedded clusters at their earliest stages of formation can provide insight into the spatial and temporal distribution of the stars and thus probe different star cluster formation models. }
     {We present near-infrared imaging of an 8\arcmin$\times$13\arcmin\ (5.4pc$\times$8.7pc) region around the massive infalling clump G286.21+0.17 { (also known as BYF73)}. The stellar content across the field is determined and photometry is derived in order to { obtain} stellar parameters for the cluster members.} 
     { We find evidence for some sub-structure (on scales less than a pc diameter) within the region with apparently at least three different sub-clusters associated with the molecular clump based on differences in extinction and disk fractions. At the center of the  clump we identify a deeply embedded { sub-cluster}. Near-infrared excess is detected for { 39-44\% in  the two sub-clusters associated with molecular material and 27\%\ for the exposed cluster. Using the disk excess as a proxy for age this suggests the clusters are very young.}}
   { The current total stellar mass is estimated to be at least 200 M$_\odot$.}
     { The molecular core hosts a rich population of pre-main sequence stars. There is evidence for multiple events of star formation both in terms of the spatial distribution within the star forming region and possibly from the disk frequency. }
\end{abstract}

\section{Introduction}
{ The process of star cluster formation remains poorly understood. }
How it proceeds through the collapse of a molecular clump, the time scale of the collapse, and whether it is a monolithic collapse or a collection of mergers of sub-clusters is still debated { \citep[][see \citet{longmore14} for a review]{elmegreen00,tan06,elmegreen2007,nakamurali,hartmannburkert}. 
Numerical simulations show that self-gravitating, turbulent,  and magnetized gas has a low star formation efficiency per free fall time \citep[e.g.][]{krumholz05,padoan11,padoan14}. 
However, if turbulence has dissipated, efficient star formation (of 30\% or more) can happen on as little as one free fall time \citep[e.g.][]{elmegreen00}. 
Most massive clusters known are already gas free such that star formation has stopped and thus in this context are relatively old. }

To be able to distinguish the different formation models one has to observe clusters during their formation. 
 Although there are massive pre-cluster candidates, e.g. G0.253$+$0.016 that may form a massive Galactic Center cluster in the future \citep{longmore12} very active star formation has not yet begun within them. 
{ The CHaMP survey \citep{barnes05} is as unbiased survey of $20\times3$ degrees of the Galactic plane to characterize the star formation \citep[see also][]{barnes11}.  One of the first discoveries out of the survey was the identification of a massive clump with a very large infall \citep{barnes10}. 
The massive clump G286.21+0.17 is located  in the Carina star forming complex and was identified to have a  molecular infall rate of ($3.4\times10^{-2}\rm{M_\odot yr^{-1}}$), indicating that star formation is still ongoing \citep[][where the core is known as BYF73]{barnes10}.}
They suggested a molecular mass of up to 2$\times$10$^4$ M$_\odot$ assuming a distance of 2.5 kpc and a luminosity of 2-3$\times10^4$ L$_\odot$ \citep[see also][]{verma94}. 
We adopt a distance of 2.3 kpc for the region based on its association near the Carina complex, see \citet{smithbrooks} for a discussion of the distance.

That star formation is ongoing was indicated already from IRAS observations and subsequent balloon observations \citep{verma94} but higher resolution observations were necessary to confirm a deeply embedded cluster in the center of the clump observable at 2.2 $\mu$m. 
\citet{ohlendorf} confirmed the presence of the cluster and presented deeply embedded sources detected in the mid-infrared with {\it Spitzer} IRAC. However the sensitivity, spatial resolution, and wavelength coverage in the previous studies complicated a characterization of the stellar population using current mid-infrared capabilities.

Here we present deep near-infrared VLT/HAWK-I \filter{$JHK_s$} imaging of G286.21+0.17 and its surrounding environment.
The main goal is to characterize the stellar content, in terms of object masses and disk fractions, of the molecular clump and of nearby possibly interacting star-forming regions within the Carina molecular cloud. 

The paper is organized as follows. 
Section 2 describes the near-infrared   data and their reduction. 
In Section 3 the basic results from the color-magnitude and color-color diagrams are presented and discussed.
The extinction map of the region based on the near-infrared photometry is presented.
 The derived disk fraction for the different sub clusters and a lower estimates for the   total mass of the star forming regions are presented. 
Further the spatial distribution within G286.21+0.17 is discussed. 
Finally we conclude in Section 4.

\section{Data}
The near-infrared HAWK-I data and their reduction are presented. The photometry is derived and converted into the 2MASS system.  The source counts are discussed and the depth of the data is derived. 

\subsection{Data reduction}
\filter{JHK$_S$} observations were obtained with HAWK-I on the VLT UT4 under program IDs 087.D-0630 and 089.D-0723 { over the period 2011-2013.} 
HAWK-I has a field of view of 7.5\arcmin\ with a pixel scale of 0.106\arcsec/pixel. 
The seeing during the observations was 0.4\arcsec-0.6\arcsec (920-1380AU). 
A mosaic of 8\arcmin$\times$13\arcmin\ was observed where the mosaicking was performed by offsets along strips to compensate for the gap between each of the four detectors and to obtain sky frames using a running median stacking of the science frames, using the adjacent frames for the sky { using {\tt imcombine} within the {\tt iraf} environment to create the sky frames}.
Each frame was obtained with an individual exposure time of 10 seconds with 6 repeats before the telescope was slewed 3\arcmin . 

\begin{figure*}
   \centering
   \includegraphics[width=16cm]{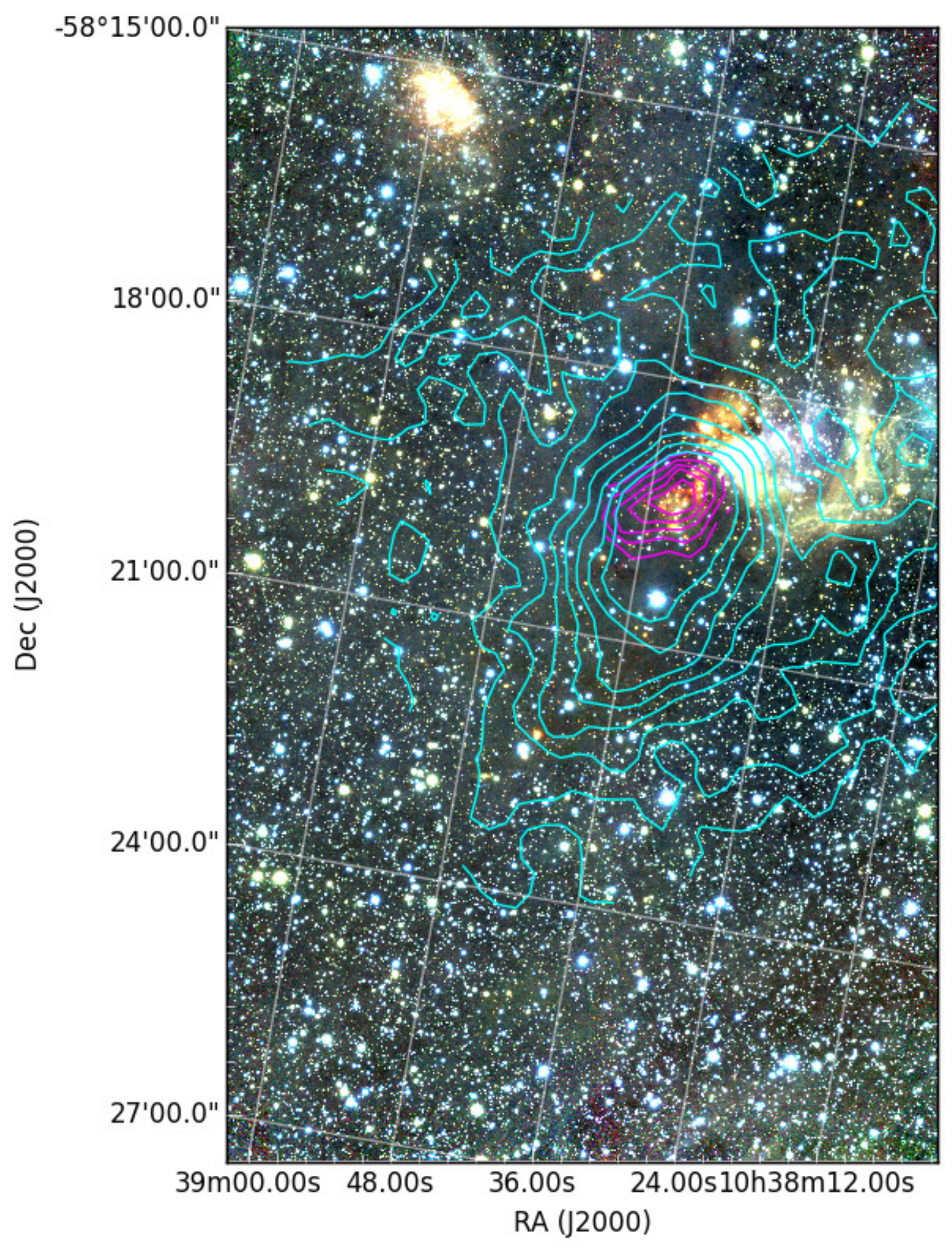}
   \caption{\filter{JHKs} (blue, green, red, respectively) color mosaic of the region around G286.21+0.17.
     The field of view is 5.5\arcmin$\times$12.7\arcmin , corresponding to 3.7$\times$8.5 pc for a distance of 2.3 kpc. 
Shown as magenta contours is the \hcob\ zero moment contours from 0.3 to 1.1 $\mathrm{K~km/s}$ in steps of 0.2 $\mathrm{K~km/s}$ and in cyan the \hcoa\ zero moment contours from 0.6 to 3.8 $\mathrm{K~km/s}$ in steps of 0.4 $\mathrm{K~km/s}$. }
              \label{overview}%
    \end{figure*}

{ Cosmic rays were identified in each frame using {\tt lacosmic} \citep{lacosmic} and were masked in the further data reduction. 
Each frame was corrected for geometric distortion based on stars in common with the 2MASS catalog { using {\tt geomap} and {\tt geotran} in the {\tt iraf} environment.} 
A third order polynomial fit based on more than 100 stars per frame was performed resulting in an RMS of 1 pixel. 
}
The frames for each filter were then median combined. 
The total exposure times across the molecular clump and star clusters were 6000 seconds in \filter{J}, 1500 seconds in \filter{H}, and 1500 second in \filter{Ks}, respectively. 
A color-representation of the observed region is shown in Fig.~\ref{overview}. 

\subsection{Source detection, calibration and photometry}
The region is located in the Galactic plane and is strongly affected by extinction both by the foreground and from the molecular clump.
This, together with the intrinsic red colors of pre-main sequence stars, makes the young population easier to identify in the \filter{Ks} band than at \filter{J} or \filter{H} despite the shorter exposure time. 
Point sources were therefore identified in the \filter{Ks} band.
The source detection was performed using {\tt daophot} within the {\tt iraf} environment, where a  detection threshold  4$\sigma$ above the background was used. 
The source list was then used for photometry in all three bands. 
Photometry was performed in a 5 pixel radius aperture (0.5\arcsec, corresponding to the average seeing) and a local sky determined in an annulus between 30 and 35 pixels (3-3.5\arcsec).

\begin{figure}
   \centering
  \includegraphics[width=8cm]{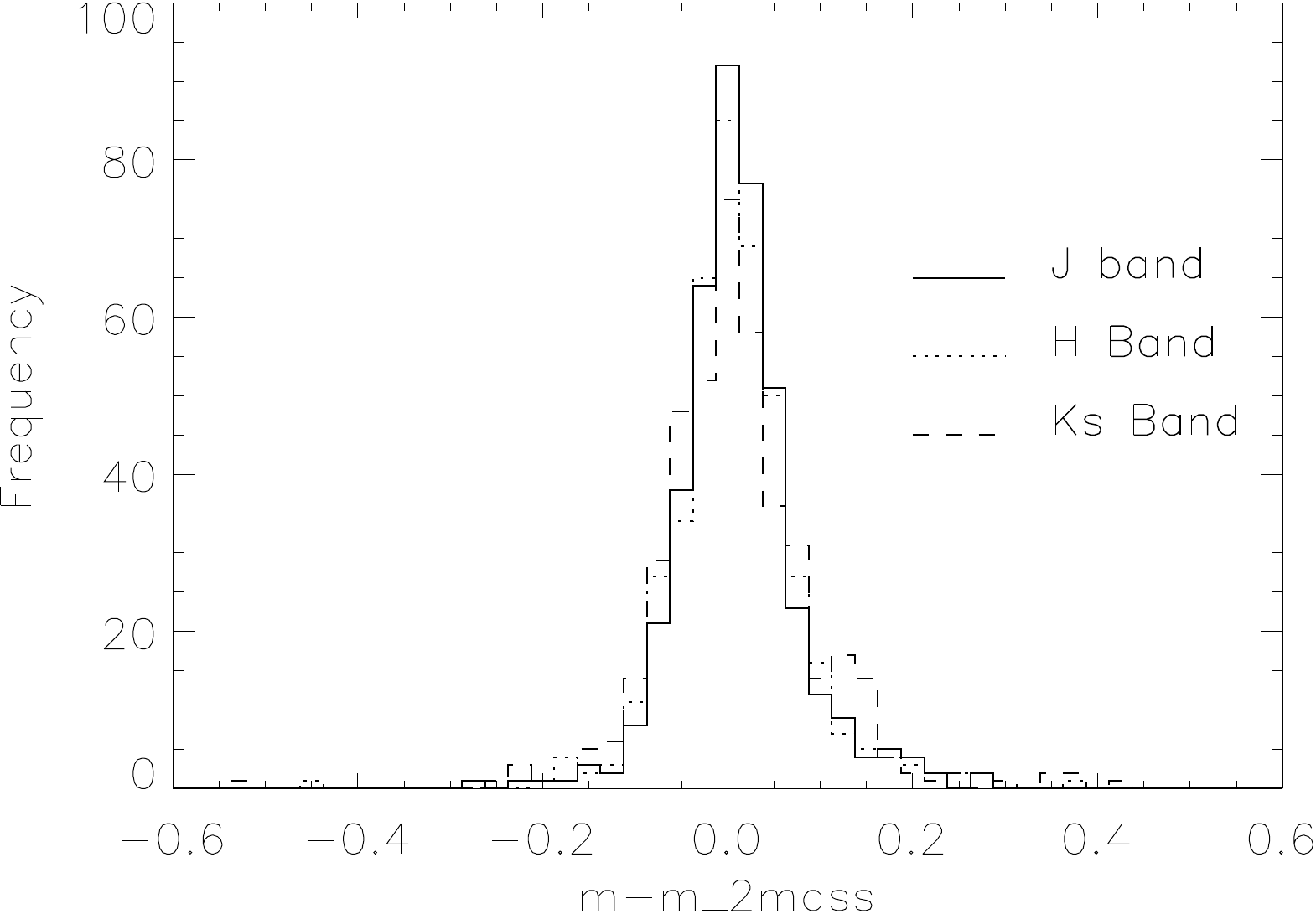}
   \caption{Histogram of the magnitude differences for each filter between the 2MASS photometry and HAWK-I photometry after the HAWK-I photometry was transformed into the 2MASS system.}
              \label{comp_2mass}
    \end{figure}

The photometry was converted into the 2MASS system utilizing  stars in common with the 2MASS catalog and not saturated in the HAWK-I data. 
Stars in common between this survey and 2MASS fainter than \mbox{\it $\mathrm{K_{2mass}}$}$=13$ mag were used. 
Zero points and first order color terms were determined for each filter. 
The comparison with the 2MASS photometry is shown in Fig.~\ref{comp_2mass} for sources with photometric errors less than 0.1 mag in the 2MASS catalog (421 sources).  
The best-fit Gaussian to each histogram are centered at 0 with a dispersion of 0.04-0.06 mag depending on the filter. 
The spread is expected given the large differences in spatial resolution and sensitivity between the two data sets.  

\begin{figure}
   \centering
  \includegraphics[width=8cm]{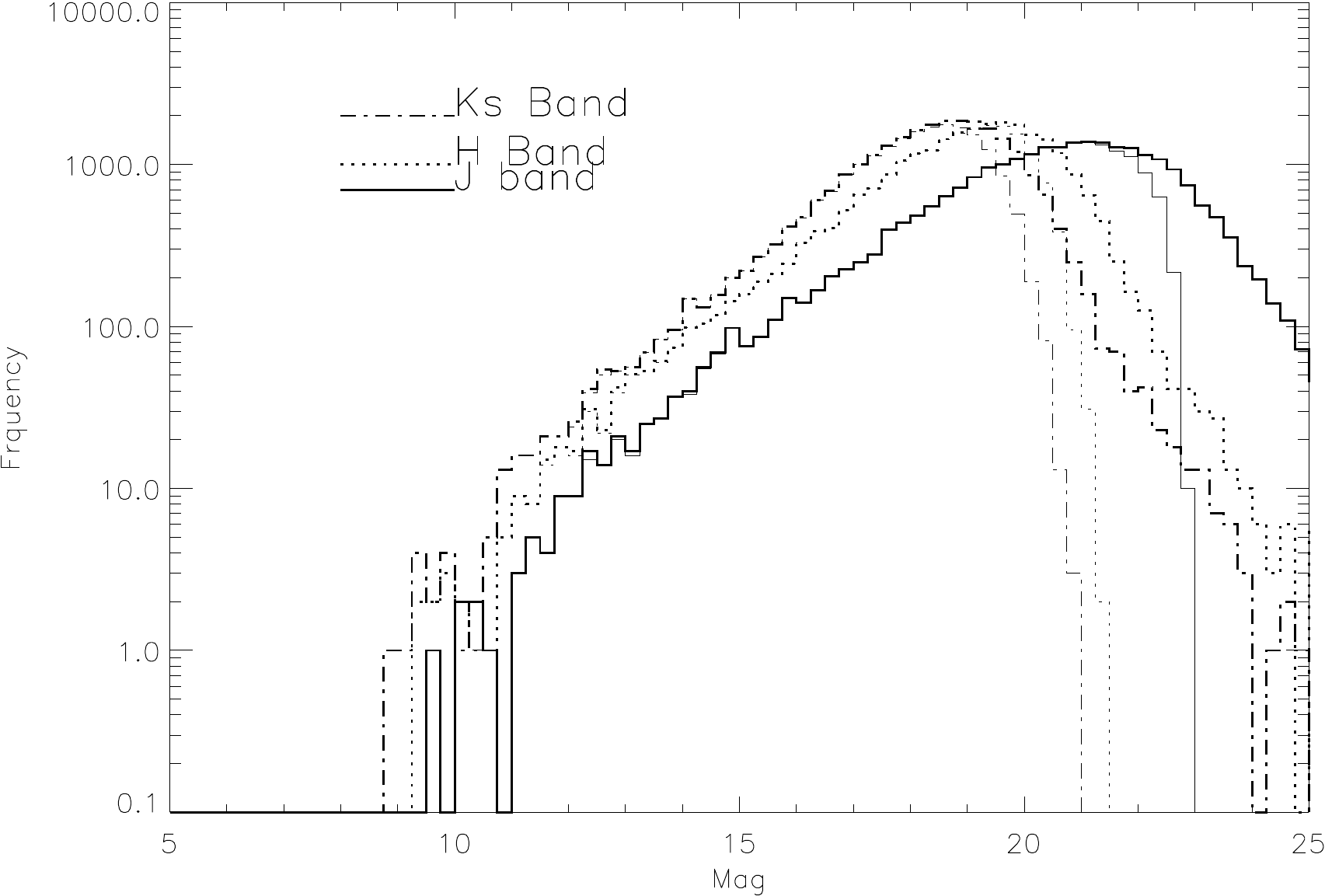}
   \caption{Number counts as a function of magnitude for the three filters. For each filter the thick line represents all detections, while the thin line is for the restriction of a magnitude error of 0.1 mag in each band. }
              \label{numbercounts}
    \end{figure}

A total of 55129 sources were detected in the \filter{Ks} band mosaic.
Using that photometry list on the \filter{J} and \filter{H} band mosaic and only accepting sources with a centroid position that agrees within 1 pixel in x and y, a list of 25875 sources have been identified. 
Further restricting the list to objects having a photometric  error less than 0.1 mag in each band results in a total of 20989 objects. 
The number counts as a function of magnitude is shown in Fig.~\ref{numbercounts}.  
Sources down to \filter{Ks}=21 mag are detected with a formal error of 0.1 mag. 
The source counts deviate from a power-law at \filter{Ks}$\approx$ 18.5 mag and \filter{J}$\approx$ 21 mag, which indicates incompleteness at fainter magnitudes.

\subsection{Completeness} 
We have performed artificial star experiments to test the depth of the data.
With the spatial resolution and the general source surface density it is expected that the background noise from the Earths atmosphere and ionized gas within the cluster region are the main limitations for detection objects. 

The tests follow a similar method as  e.g. \citet{andersen09,andersen15}. 
Briefly, a point spread function (PSF) is created from isolated bright non-saturated stars across the field.
Artificial stars are then placed in the frames using this PSF and magnitudes similar to those observed.
The number of artificial stars added is limited to 10\%\ of the observed stars in the region analyzed to avoid changing the crowding characteristics.
The artificial stars follow a luminosity function similar to that of the stars observed.
The stars are placed at a random position using the PSF before aperture photometry is performed for all objects. 

\begin{figure}
   \centering
  \includegraphics[width=8cm]{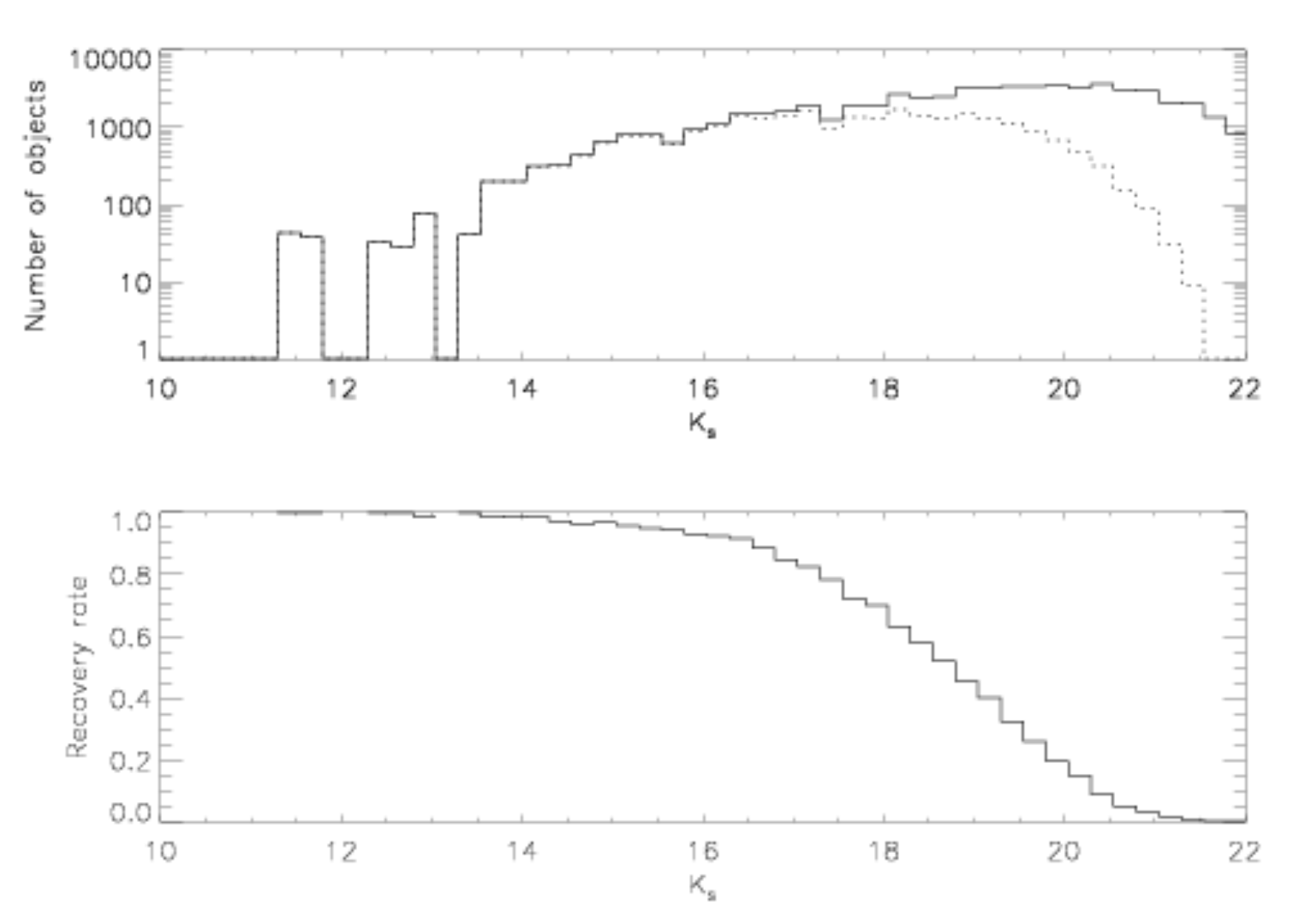}
  \caption{Top: Histograms of the \filter{Ks} band magnitudes of all the artificial stars placed in the frame (solid lined histogram) and of the retrieved stars (dotted histogram). Lower: The ratio of the two histograms as a function of \filter{Ks} band magnitude.}
  \label{arti}
    \end{figure}

An artificial star was considered recovered if the retrieved position was less than 0.5 pixels (0.05\arcsec ) from the input position and the retrieved magnitude is less than 0.1 mag from the input value.
Fig.~\ref{arti} shows the recovery frequency as a function of magnitude. 
    The photometry across the cluster regions is { more than 80\%}  complete for sources brighter than \filter{Ks}$=$17 and is 50\%\ or better  complete for sources down to \filter{Ks}$=$19. 

    To estimate the total cluster content and the total mass of the (sub) clusters we utilize the completeness correction to take the missing stars into account.
    In practice this is performed by weighing each object detected with the completeness correction.
    A { function with the shape of the Fermi function, $f(m)=1/(1+e^{(m-m_0)/m1})$, was fitted to the completeness as a function of magnitude and the completeness correction for each star was then determined from the fit. 
    
\section{Results and Analysis} 
We present  the immediate results from the imaging survey. 
The color-magnitude and color-color diagrams are utilized to characterize the general stellar population within the field of view and to identify several sub-clusters. 
The sub-cluster content is then described in more detail and their relation to the surrounding molecular cloud is discussed.

\subsection{Overview of the region}
Fig.~\ref{overview} shows the near-infrared observations together with the \hcob\ contours obtained from the observations discussed in \citet{barnes10}.
The clump G286.21+0.17 is seen as the molecular emission peak at the middle of the image.
\citet{barnes10,ohlendorf} noted the presence of embedded sources within the clump but it is only with the sensitivity and spatial resolution in the near-infrared that the relative richness of the cluster is evident.
Most of the sources are heavily obscured and only { detected} in the \filter{Ks} band.

To the north-west of G286.21+0.17 is located more exposed star formation as noted in \citet{barnes10}.
{ The strong diffuse emission in a shell-like structure suggests there to be an \HII\ region present, which is confirmed by the $\mathrm{Br_\gamma}$ emission seen by \citet{barnes10}.}
This region is also more strongly populated by stars and most are less affected by extinction compared to the deeply embedded cluster.

The combination of the \hcoa\ and \hcob\ emission confirms the conclusions based on the stellar colors.
There is little molecular material associated with the apparently exposed \HII\ region whereas the red stellar cluster is located at the peak of the \hcob\ emission and thus heavily embedded.
There appears to be substantial lower column density material extended north-south along the star forming regions.
Curiously there appear to be few red stars associated with this structure.

At the northern edge of the field of view is located a large diffuse emission region. 
In its outskirts is further seen a thin bow structure both to the north- and south-west. This object is located at a projected distance of 3.7 pc { from G286.21+0.17} and is most likely not directly related to the star formation event discussed here but a part of the greater Carina molecular cloud.

Fig.~\ref{zoom} shows the region surrounding G286.21+0.17.
Three cluster regions are marked, one centered on the stars in the molecular clump (region 1), one centered on the diffuse emission region (region 3), and one centered on the interface region of the two (region 2).
Further, the whole complex is also studied as one  to show the presence of sub-structure justifying the separation into several groups. 
A control field was further chosen to be able to quantify the fore- and back-ground contamination. The location of the control field, { shown in Fig.~\ref{zoom} as a yellow circle} was chosen to be East of the cluster at similar Galactic latitude to probe a similar field population.
{ The cluster regions are selected by hand based on the different characteristics of the different locations, mainly in terms of the molecular material associated with each cluster.  }

    \begin{figure}
   \centering
  \includegraphics[width=9cm]{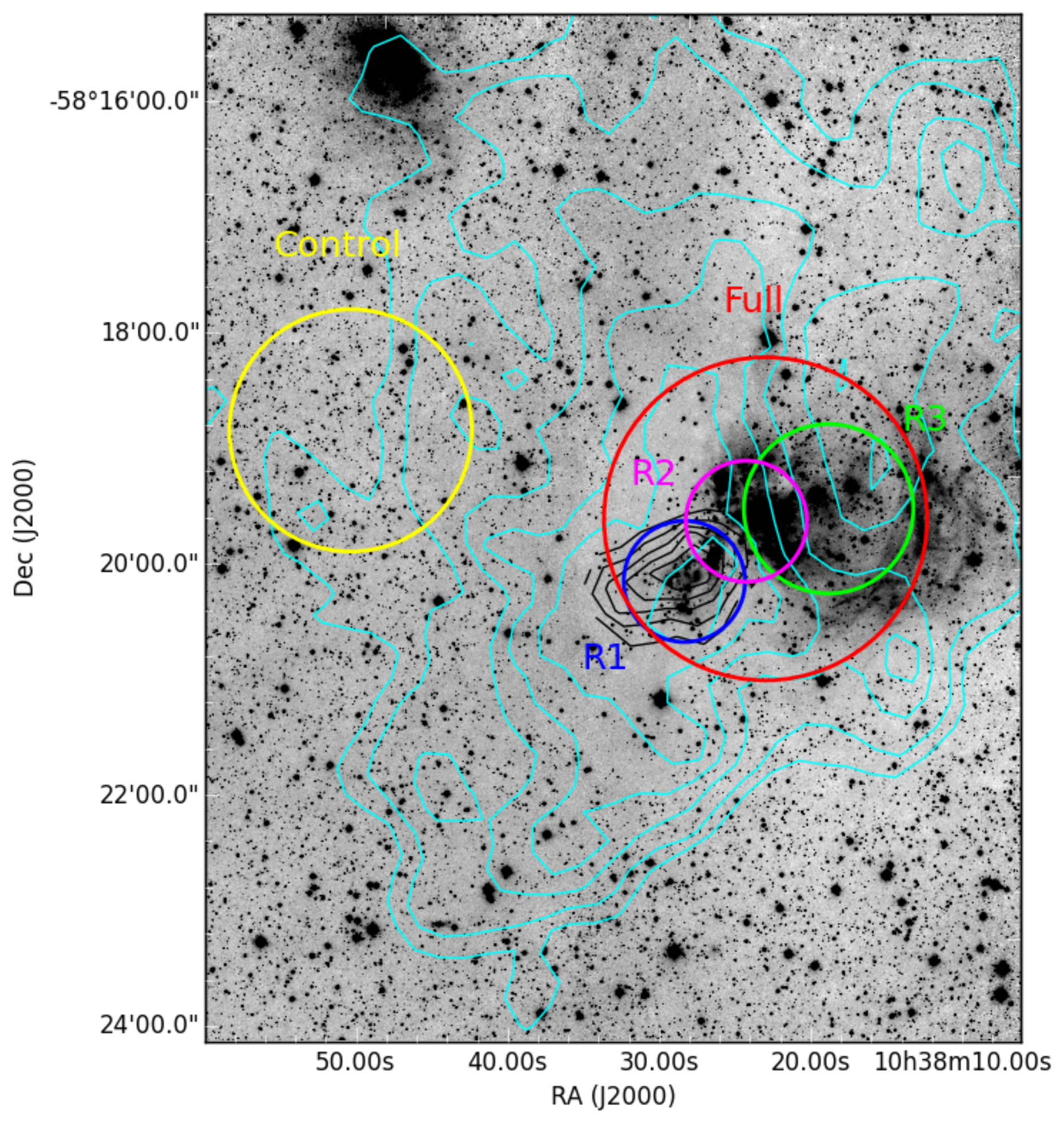}
    \caption{\filter{Ks} band image showing the region surrounding G286.21+0.17. Shown are the \hcob\ contours (black) from 0.3 $\mathrm{K km/s}$ to 1.1 $\mathrm{K km/s}$ in steps of 0.2 Kkm/s and the extinction map contours in steps of A$_\mathrm{V}$=6,8,12,16,20,24 (cyan). 
North is up, east to the left.  
The location of the selected cluster regions are shown as circles and marked. The control field selected is shown in yellow.  }
              \label{zoom}
    \end{figure}

\subsection{Color-magnitude diagrams}

The \filter{J-H} versus \filter{J} color-magnitude diagrams for the cluster regions and the control field are shown in Fig.~\ref{CMD_all}. 
   \begin{figure}
   \centering
  \includegraphics[width=8cm]{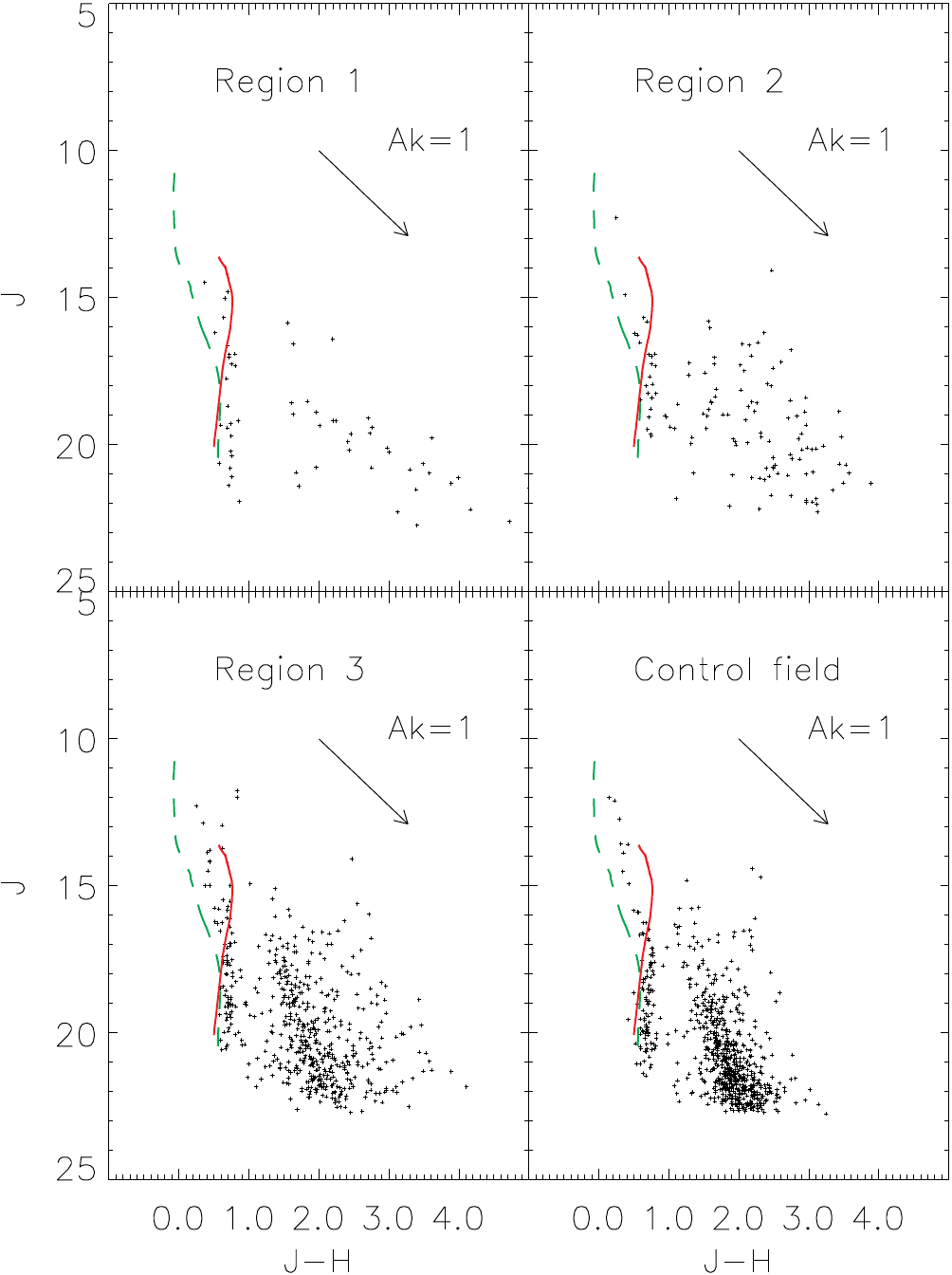}
  \caption{\filter{J-H} versus \filter{J} color-magnitude diagrams for the three selected sub-regions shown in Fig.~\ref{zoom} and for the control field.
    Overplotted are a reddening line and the main sequence \citep{marigo} (green) and a \citet{baraffe} 1 Myr isochrone (red), both shifted by a distance of 2.3 kpc. }
  \label{CMD_all}
    \end{figure}
The control field shows two distinct populations, one at \filter{J-H}$\approx$0.6 and another at \filter{J-H}$\approx$ 2.    
Comparing the  main sequence in the color-magnitude diagrams with the bluer stars in the control field suggests that these are  in the Galactic disk and slightly reddened, consistent with being the field population on the foreground to the clump.
This would suggest a foreground extinction of a few tenths in A$_{Ks}$, consistent with a typical extinction in the Galactic plane of A$_\mathrm{V}\approx1.8$/kpc, \citep{whittet}. 

The redder population is likely to consist of a background population reddened by the line-of-sight material in the spiral arm hosting the Carina Nebula.
There might also be slight added field population from the general extended star formation in the Carina Complex.
Some evidence for this is provided in the color-color diagram of the control field shown below. { Thus, for the statistical subtraction of field objects it is assumed this population is constant across the field of view, which is small compared to the size of the Carina star forming complex.} 

The color-magnitude diagrams for the sub-clusters and the control region show very distinct morphologies. 
In all three regions the foreground population clearly stands out at \filter{J-H}$\sim$0.6 and a \filter{J} magnitude range of 15-20.  
There is a large range of extinction without any clear concentration. 
Based on the amount of extinction determined from the mm observations \citep{barnes10} the detection of background field stars is expected to be modest. 

Region 2, adjacent to G286.21+0.17, displays a more complicated behavior.
The foreground population is still clearly visible.
The region appears more rich than region 1.
This, however, may be due to the much larger extinction towards region 1 that partly obscures members in region 1.
There is a large range of reddening for this region as well suggesting that the stellar population is still partly embedded.

The exposed HII region (region 3)  has a morphology very similar to the control field but there is a slight overabundance of sources with  a color of \filter{J-H}$\approx 2$ and \filter{J}$\approx 17$, which is the bright end of the cluster population.

\subsection{Field star contamination} 
It is expected that a large fraction of the stars detected are field objects and part of a more dispersed population of previous star forming events in the Carina molecular cloud. 
Both will affect the derived cluster parameters if not taken into account. 

Part of the foreground population is well separated from the expected cluster content.
The red population seems to overlap with the cluster content in the color-magnitude diagram and thus needs to be accounted for. 
We have statistically subtracted the field contamination using the same technique as in \citet{gennaro,andersen15}.
The algorithm calculates the density of objects in the color-color diagram of   both the designated field region and the cluster region taking the photometric errors into account. { Both density diagrams are then scaled by the surface area on the sky to enable a comparison of the object density per unit area.} 
The relative density of objects in the two diagrams is then  used to assign a probability that an object in the cluster is a field object.
{ Membership is then determined by comparing the probability for each object with a randomly drawn number between zero and unity. If the number is greater than the probability the object is deemed a field object.} 
This approach worked very well for  Westerlund 1 \citep{gennaro,andersen15} due to the high density of objects and limited range of colors providing well sampled probability density functions.

Note that the observed region is within the Carina molecular cloud and some contamination of other star forming events is likely. 
{ Indeed a small disk fraction is observed in the control field, as shown below. }

\subsection{Color-color diagram and disk fractions} 
The near-infrared color-color diagram is  an effective tool to identify objects with a warm circumstellar disk \citep[e.g.][]{meyer97}. 
The heating of the disk is responsible for an excess at longer wavelengths separating the objects in the color-color diagram from objects without a disk with only the stellar photosphere present.  
Fig.~\ref{CCD_all} shows the \filter{H-K} versus \filter{J-H} diagrams for the described sub-regions and the control field. 
\begin{figure}
   \centering
  \includegraphics[width=8cm]{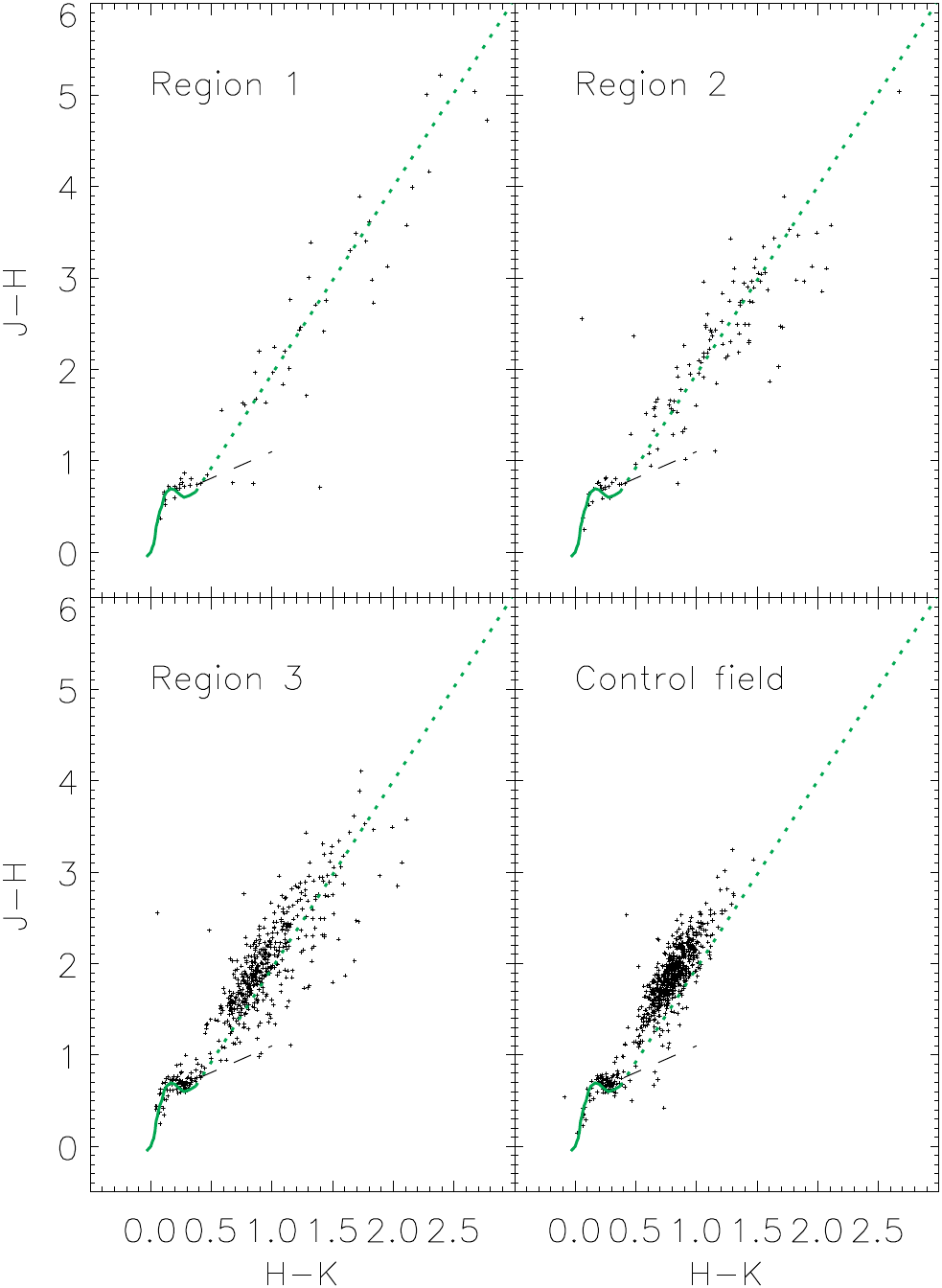}
   \caption{\filter{H-K} versus \filter{J-H} color-color diagram for the three cluster regions and the control field shown in Fig.~\ref{zoom} for all objects with photometric errors less than 0.1 mag in each band. Overplotted are the main sequence dwarf sequence \citep{besselbrett}, { a the reddening vector extending from an M6 spectral type and the T-tauri locus from \citet{meyer97}}. }
              \label{CCD_all}
    \end{figure}
The foreground population follows the colors of the main sequence as expected but with a slight shift ( a few tenths magnitudes in color) due to the foreground extinction. 
As is the case for the color-magnitude diagram, the surveyed region as a whole shows a large spread in reddening but with the majority of the objects having colors consistent with being reddened field stars or reddened background giants. 

Objects located to the right of the reddening vector from the M6 main sequence colors are typically assumed to be objects with an excess due to a circumstellar disk.
{ Further, the objects have to be above the de-reddened T-Tauri locus to be considered T-Tauri star candidates. } 
Even in the control field there are a few objects showing evidence for a disk excess. 
This is not surprising given the extended star formation in the Carina molecular cloud. 

We thus determined the disk fraction based on the ratio of objects to the left and to the right of the reddening vector from a non-reddened M6 main sequence star { and above the T-Tauri locus} for the three cluster fields and the control field, which is the typical method from near-infrared color-magnitude diagrams.

In addition, all objects with a color \filter{H}-\filter{Ks}$<0.5$ have been ignored since they are foreground objects. 
Table~\ref{disk_frac} presents the disk fractions for the selected regions within the observed field.
{ Although the disk fraction in Region 3 is relatively high it is still lower than that of Regions 1 and 2. Adopting the disk fraction as a proxy for (relative) age this would suggest an age difference between the regions, supported by the difference in how embedded the regions are; Region 3 is almost exposed already.} 
{ In addition to the full sample we have further created a magnitude and extinction limited sample for each of the cluster in order to probe a similar range of stellar masses for the different regions.
We have set the limiting magnitude and extinction as \filter{J}$<$19.5 and $\mathrm{A_K}=2$ as a compromise between depth and being able to reach a sample probing a reasonable range of extinction. }
\begin{table}
\caption{The disk fraction for the sub-regions shown in Fig.~\ref{zoom}}
\begin{tabular}{lllll}
Region & Objects & Ratio & Objects lim & Ratio\\
\hline
Region 1 &31&63$\pm$7\%&12&39$\pm$16\%\\
Region 2 &85&49$\pm$5\%&33&44$\pm$7\%\\
Region 3 &327&29$\pm$2\%&117&27$\pm$3\%\\
Control &626&5$\pm$1\%&99&3$\pm$2\%\\
\end{tabular}
\label{disk_frac}
\tablecomments{After field subtraction and the corresponding fraction for the control field. The disk fractions for the magnitude and extinction limited sample are also given. Error bars represent the uncertainty from counting statistics and the field subtraction.}
\end{table}
 
The disk fraction for all three cluster regions is high compared with other star forming regions \citep{hillenbrand} and would suggest ages of around 1 Myr. 
The determined disk fraction appears marginally higher for regions 1 and 2 than for region 3 suggesting they are younger than the exposed \HII\ region.
{ However, as seen in Fig.~\ref{CMD_all} there is a  substantially larger spread in determined extinction values for region 1 in particular and partly for region 2. 
Objects are only detected in the \filter{Ks} band and are thus not included in the samples since several bands are necessary to identify a disk and the higher extinction would exclude them from the current extinction limited sample. 
 The majority of stars in region 1 are too heavily embedded to be detected in the VLT \filter{J} band. Deeper HST observations are expected to reach deeper embedded objects also at short wavelengths. }  
Their relative youth is in general agreement with the two sub-clusters being more deeply embedded and thus not have had time to clear the surrounding material.

\subsection{Near-infrared extinction map and the molecular gas mass} 
The near-infrared photometry has been used to construct an extinction map of the region which is shown in Fig.~\ref{zoom}.
 The method follows the approach in \citet{kainulainen}, { although limited to the near-infrared observations} and is usable to extinction values up to $\mathrm{A_V}\approx20$ at which point the surface density of stars detected in the \filter{J} band becomes too low to obtain a reasonable spatial sampling. { For higher extinction regions the \hcob\ map provide a good estimate.} 
The background stellar density allows a resolution of the extinction map of 21\arcsec , corresponding to a physical scale of 0.27 pc for a distance of 2.3 kpc. % and twice the resolution of the \hco\ maps. 

Due to the distance of 2.3 kpc there is expected to be unrelated material along the line of sight which would overestimate the extinction associated with G286.21+0.17. 
Based on the location of the expected foreground stars in the color-magnitude diagrams in Fig.~\ref{CCD_all}, the typical color is \filter{J}-\filter{H}$\approx 0.7$. 
A typical color for late-type stars is \filter{J}-\filter{H}$\sim$ 0.5-0.6. 
Using the extinction law of \citet{calzetti} this corresponds to A$_\mathrm{V}\sim$1.5 which is subtracted from the extinction measures. 
Changes in this adopted foreground extinction of 1 magnitude changes the total mass estimate by 10\%.

The gas mass determined from the extinction map is 1600 M$_\odot$ assuming A$_\mathrm{V}$=1.87$\times10^{21}$\ N$_\mathrm{H}$ \citep{bohlin} assuming $\mathrm{R_V}=3.1$  after foreground subtraction. 
This is a strict lower limit since it only accounts completely for the mass in regions where the extinction can be well determined.
High column density gas tracers or long wavelength observations of the dust seen in emission can be used for the densest parts.
The analysis by \citet{barnes10} suggested a gas mass of at least $10^4$ M$_\odot$ for the central parts of the core where the extinction map is not sensitive { and \citet{barnes16} find a similar mass but for the extended cloud using $\mathrm{^{12}CO}$}.

\subsection{Cluster sub-structure}
A search for sub-structure in star clusters can provide information on its evolutionary state. 
One way to parameterise the structure is the Q value introduced by \citet{cartwright}. 
The method utilizes the Minimum Spanning Tree (MST) and the  ratio of the normalized mean length in the branches of the tree  to the ratio of the mean separation to the cluster radius, $Q=\frac{\overline{m}}{\overline{s}}$ is a measure of the structure within the region examined. 
As illustrated in \citet{cartwright} a Q value larger than 0.8 indicates a centrally concentrated whereas a smaller ratio indicate a sub-structured (fractal) distribution. 

We have calculated the Q parameter for all the objects deemed members from the statistical background analysis for the three sub-clusters and for the integrated region shown in Fig.~\ref{zoom} as the red circle.
{ In all cases we used the magnitude and extinction limited samples to limit biases due to different depth and extinction values.} 
The resulting values are indicated in Fig.~\ref{Q} with the same color coding as in Fig.~\ref{zoom} for each region.
{ Further is shown the distribution of Q values for samples of 20 objects randomly drawn from the control field.} 

The width of the distribution of the random samples from the control field suggests that the sample sizes are relatively small to determine substructure or central concentrations.
Region 1 { shows some evidence for sub-structure based on the low Q value,} regions 2 and 3 as well as the whole region show some sign of central concentration.
{ However, the extinction limited sample for region 1 is still based on 7 objects and would benefit from a larger sample.} 
The whole region is dominated by region 3 in numbers which may complicate the search for any substructure. 
\begin{figure}
   \centering
  \includegraphics[width=8cm]{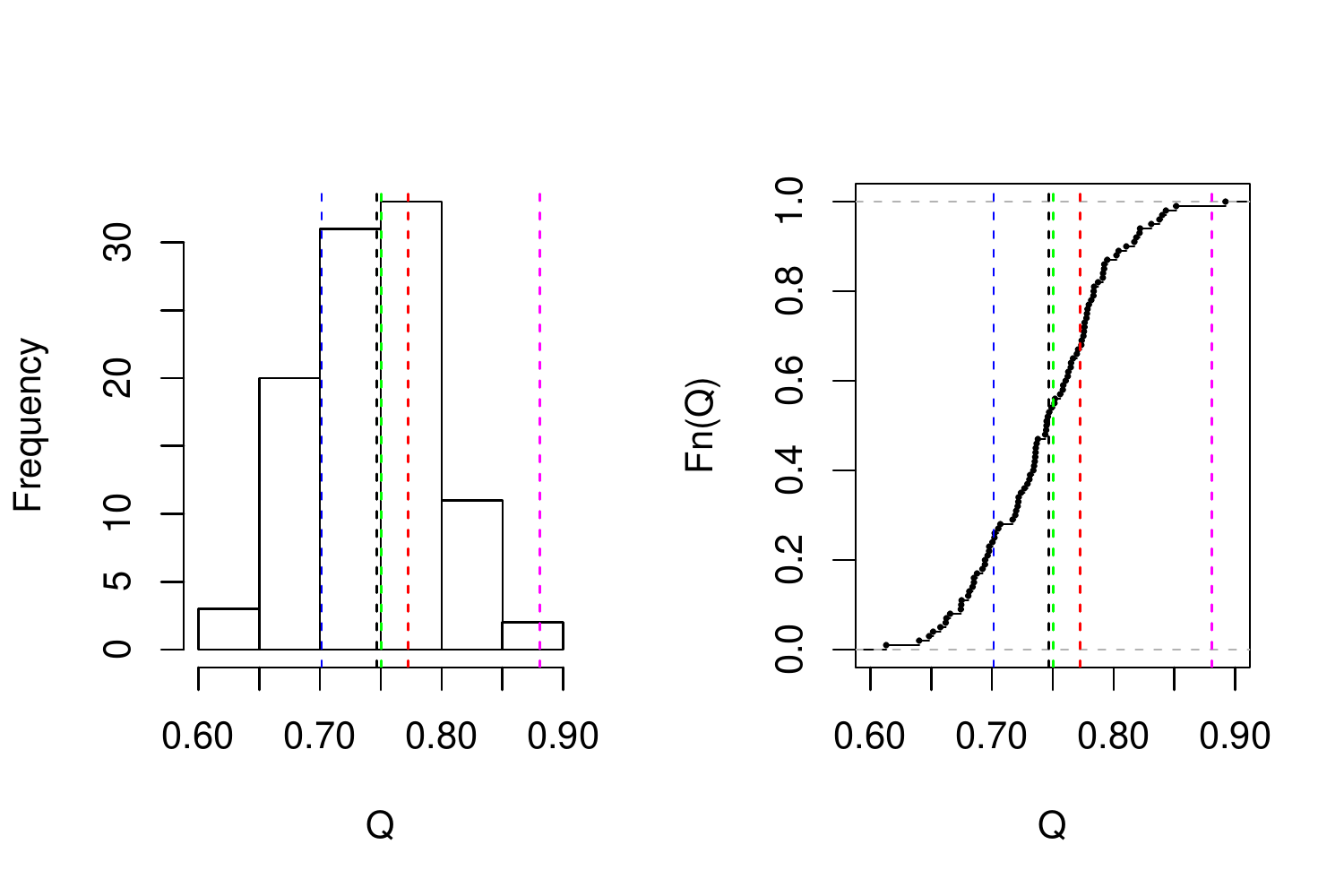}
   \caption{The distribution of Q values derived for samples of 20 stars drawn randomly from the control field. The red line indicates the value determined for G286.21+0.17 and the blue line the value for the western cluster respectively. The black line represents the Q value for the control field. }
              \label{Q}
    \end{figure}
   A larger sample is necessary to distinguish, for example through deep space based \filter{J} and \filter{H} band imaging. 

{    The Q values determined for the sub-clusters are similar to the Q valued determined in a large set of more evolved clusters by \citet{jaehnig}. 
The Q values for the regions are 0.70 for Region 1, 0.88 for Region 2, 0.75 for Region 3, and 0.77 for the full region. The Q value for Region 2 is high but similar values do occur in the sampling the control field and the Region 2 value is only one of the four regions we calculate Q for.}

\subsection{Cluster mass and age}
In principle the age of the cluster populations can be determined from the color-magnitude diagrams. 
The time it takes a star to arrive on the main sequence is dependent on mass and the turn-on point where the pre-main sequence population merges with the main sequence is thus an age indicator once the distance to the cluster is known. 
Differential extinction complicates this analysis for relatively sparse clusters since it is not clear where the objects should be dereddened to, the main sequence or the pre-main sequence once an age is adopted. 
If the main sequence is richly populated the distribution of extinction can be estimated and the average location of the pre-main sequence objects in the CMDs can be found. 
Here we assume an age of 1 Myr and adopt the isochrones from  \citep{tognelli} and \citet{baraffe} converted into the 2MASS system using the TADA software \citep{dario_tada} for a direct comparison with the observed magnitudes.
{ An age of 1 Myr is based on the generally high disk fraction and the association with the molecular material which suggest youth and is consistent with an age of 1 Myr.
  An actual younger age would result in an overestimate of the pre-main sequence stellar content and vice versa for an underestimate.}
For objects above 1.4 M$_\odot$ we use the \citet{tognelli} isochrone and below  \citet{baraffe} since their evolutionary tracks extend to lower masses.
\citet{scandariato} showed that the model magnitudes of pre-main sequence stars are differ systematically  when compared to observations for late spectral types \citep[e.g.][]{scandariato}, in particular in the \filter{H} band due to the strong water band absorption. 
A correction to the model isochrones as a function of effective temperature was tabulated by \citet{scandariato} for the Orion Nebula Cluster, which has an age of 1.5-3.5 Myr \citep{reggiani,dario12}, for objects in the spectral class range K6-M8.5, covering the PMS stars in our sample.  
A linear interpolation for the color correction as a function of effective temperature was performed based on the effective temperature from the evolutionary models.
This  provides the predicted magnitudes from the luminosity and effective temperatures that can then be compared with the observed magnitudes and colors.

{ The total mass estimate for each sub-region and the whole cluster are provided in Table~\ref{masses}. 
All the mass estimates are to be considered lower limits since it is expected that stars will remain undetected due to extinction. 
This is particularly severe for region 1 but can also affect the other regions.Further, for a more direct comparison between the different regions a lower mass limit of 0.15 M$_\odot$ has been adopted.} 
 
Both the directly observed mass and the corrected mass estimates are provided. 
{ There are two corrections provided to the measured mass. The first is due to incompleteness: As discussed in Section 2.3 only a fraction of the fainter stars is detected due to crowding and variable background. Thus, the mass for each star is weighted by the inverse of the completeness determined in Section 2.3. 
The other correction is due to only a fraction of the IMF is traced. We are determining the mass between 0.15 M$_\odot$ and 7 M$_\odot$ and is thus extrapolating the mass to the total mass assuming a standard log-normal IMF at low masses and a Salpeter slope above 7 M$_\odot$ up to 120 M$_\odot$.
The correction factor is $\sim$30\%.}

\begin{table}
\label{masses}
\caption{Mass estimate for each of the  sub-clusters.}
\begin{tabular}{lllll}
Region & N & Mass & Corrected mass & Extrapolated mass\\
 & & M$_\odot$ & M$_\odot$ & M$_\odot$\\
\hline
Region 1 &           25 &           24 &           25 &           33\\
Region 2 &           58 &           72 &           74 &           96\\
Region 3 &          143 &          123 &          127 &          165\\
Full &          200 &          178 &          185 &          241\\
\end{tabular}
\tablecomments{The stellar mass in the region in the mass range 0.15 M$_\odot$ to 7 M$_\odot$. The total mass extending the mass range to a  full Chabrier IMF up to 120 M$_\odot$. Note that the assigned sub-systems overlap.} 
\end{table}

{ The mass estimates take into account all detected sourced. 
However, the preferential lack of detected sources in  region 1 due to the very large extinction makes the mass estimate for this region a stronger lower limit than for the other regions. { Including sources only detected in the \filter{H} and \filter{Ks} bands would double the sample. }
However, Even in the \filter{H} band many sources invisible and only seen in in the \filter{Ks} band images. }
Further deeper short wavelength observations will address this. 

In any case, the mass of the whole region is comparable to regions known to form massive stars, e.g. the Orion Nebula Cluster \citep[e.g.][]{dario12}. 
Given the large gas infall onto region 1 it is very like this sub-cluster will increase in mass in the future.

{  That the embedded cluster is physically close could suggest interaction between them and possibly triggering of star formation in the embedded cluster. From the near-infrared photometry alone there is little to indicate if triggering is the reason for star formation in Region 1 or if it was collapsing by itself. Future work combining the current dataset with proper motion studies and the gas dynamics will address this issue.}

%______________________________________________________________

\section{Conclusions}
We have performed deep near-infrared imaging of the cluster associated with the massive molecular clump G286.21+0.17. 
The spatial resolution is 0.6\arcsec\ and the depth of the observations is \filter{J}$\sim$21 and \filter{Ks}$\sim$18.5. 
The molecular clump is shown to be associated with a rich exposed cluster as well as two smaller embedded clusters.
{ The disk fraction is determined to be high in all three regions, ranging from 27-44\% for a magnitude and extinction limited sample. }
The total molecular mass still remaining has been estimated to be as high as $10^4$ M$_\odot$ based on high column density tracers. 
We have analyzed the spatial structure using the Q parameter. 
There is evidence for sub-structure in the star forming region. The exposed region further shows evidence for being centrally condensed.

\begin{acknowledgements}
Based on observations made with ESO Telescopes at the La Silla Paranal Observatory under programme IDs 087.D-0630(A) and 089.D-0723(A).  
PJB acknowledges support from NASA/JPL contract RSA-1464327, NSF grant AST-1312597, NASA-ADAP grant NNX15AF64G, and the UF Astronomy Department.
\end{acknowledgements}


\begin{thebibliography}{}


\bibitem[Andersen et al.(2009)]{andersen09} Andersen, M., 
Zinnecker, H., Moneti, A., et al.\ 2009, \apj, 707, 1347 



\bibitem[Andersen et al.(2016) accepted]{andersen15} Andersen, M. et al. 2016, \aap\ accepted


\bibitem[Baraffe et al.(1998)]{baraffe} Baraffe, I., Chabrier, 
G., Allard, F., \& Hauschildt, P.~H.\ 1998, \aap, 337, 403 
 
{ \bibitem[Barnes et al.(2005)]{barnes05} Barnes, P.~J., Yonekura, Y., Wong, T., et al.\ 2005, IAU Symposium, 235, 247 } 

\bibitem[Barnes et al.(2010)]{barnes10} Barnes, P.~J., Yonekura, 
Y., Ryder, S.~D., et al.\ 2010, \mnras, 402, 73 

{ \bibitem[Barnes et al.(2011)]{barnes11} Barnes, P.~J., Yonekura, Y., Fukui, Y., et al.\ 2011, \apjs, 196, 12 } 

\bibitem[Barnes et al.(2016)]{barnes16} Barnes, P.~J., Hernandez, A.~K., O'Dougherty, S.~N., Schap, W.~J., III, \& Muller, E.\ 2016, \apj, 831, 67 

\bibitem[Bessell 
\& Brett(1988)]{besselbrett} Bessell, M.~S., \& Brett, J.~M.\ 1988, \pasp, 100, 1134 


\bibitem[Bohlin et al.(1978)]{bohlin} Bohlin, R.~C., Savage, 
B.~D., \& Drake, J.~F.\ 1978, \apj, 224, 132 



\bibitem[Calzetti et al.(1994)]{calzetti} Calzetti, D., Kinney, 
A.~L., \& Storchi-Bergmann, T.\ 1994, \apj, 429, 582 



\bibitem[Cartwright 
\& Whitworth(2004)]{cartwright} Cartwright, A., \& Whitworth, A.~P.\ 2004, \mnras, 348, 589 



\bibitem[Da Rio et al.(2012)]{dario12} Da Rio, N., Robberto, 
M., Hillenbrand, L.~A., Henning, T., \& Stassun, K.~G.\ 2012, \apj, 748, 14 









\bibitem[Da Rio 
\& Robberto(2012)]{dario_tada} Da Rio, N., \& Robberto, M.\ 2012, \aj, 144, 176 



\bibitem[Elmegreen(2000)]{elmegreen00} Elmegreen, B.~G.\ 2000, 
\apj, 530, 277 


\bibitem[Elmegreen(2007)]{elmegreen2007} Elmegreen, B.~G.\ 2007, 
\apj, 668, 1064 


\bibitem[Garrod et al.(2008)]{garrod} Garrod, R.~T., Weaver, 
S.~L.~W., \& Herbst, E.\ 2008, \apj, 682, 283 





\bibitem[Gennaro et al.(2011)]{gennaro} Gennaro, M., Brandner, 
W., Stolte, A., \& Henning, T.\ 2011, \mnras, 412, 2469 


\bibitem[Hartmann 
\& Burkert(2007)]{hartmannburkert} Hartmann, L., \& Burkert, A.\ 2007, \apj, 654, 988 




\bibitem[Hillenbrand(2008)]{hillenbrand} Hillenbrand, L.~A.\ 2008, 
A Decade of Extrasolar Planets around Normal Stars Proceedings of the Space 
Telescope Science Institute Symposium, held in Baltimore, Maryland May 2-5, 
2005.~ Edited by Mario Livio, Kailash Sahu and Jeff Valenti, Space 
Telescope Science Institute, Baltimore Series: Space Telescope Science 
Institute Symposium Series (No.~19) ISBN: 9780521897846 Publication date: 
June 2008, 196 pages, pp.84-105, 84 


\bibitem[Jaehnig et al.(2015)]{jaehnig} Jaehnig, K.~O., Da Rio, 
N., \& Tan, J.~C.\ 2015, \apj, 798, 126 



{ 
\bibitem[Kainulainen \& Tan(2013)]{kainulainen} Kainulainen, J., \& Tan, J.~C.\ 2013, \aap, 549, A53 
}
 



\bibitem[Kroupa(2002)]{kroupa} Kroupa, P.\ 2002, Science, 295, 
82 

\bibitem[Krumholz 
\& McKee(2005)]{krumholz05} Krumholz, M.~R., \& McKee, C.~F.\ 2005, \apj, 630, 250 




\bibitem[Lada 
\& Lada(2003)]{ladalada} Lada, C.~J., \& Lada, E.~A.\ 2003, \araa, 41, 57 


\bibitem[Longmore et al.(2012)]{longmore12} Longmore, S.~N., 
Rathborne, J., Bastian, N., et al.\ 2012, \apj, 746, 117 


\bibitem[Longmore et al.(2014)]{longmore14} Longmore, S.~N., 
Kruijssen, J.~M.~D., Bastian, N., et al.\ 2014, Protostars and Planets VI, 
291 




\bibitem[Marigo et 
al.(2008)]{marigo} Marigo, P., Girardi, L., Bressan, A., et al.\ 2008, \aap, 482, 883 


\bibitem[Meyer et al.(1997)]{meyer97} Meyer, M.~R., Calvet, N., 
\& Hillenbrand, L.~A.\ 1997, \aj, 114, 288 


\bibitem[Nakamura 
\& Li(2007)]{nakamurali} Nakamura, F., \& Li, Z.-Y.\ 2007, \apj, 662, 395 




\bibitem[Ohlendorf et 
al.(2013)]{ohlendorf} Ohlendorf, H., Preibisch, T., Gaczkowski, B., et al.\ 2013, \aap, 552, A14 


{ \bibitem[Padoan \& Nordlund(2011)]{padoan11} Padoan, P., \& Nordlund, {\AA}.\ 2011, \apjl, 741, L22 } 

{ \bibitem[Padoan et al.(2014)]{padoan14} Padoan, P., Federrath, C., Chabrier, G., et al.\ 2014, Protostars and Planets VI, 77 } 
 
\bibitem[Reggiani et 
al.(2011)]{reggiani} Reggiani, M., Robberto, M., Da Rio, N., et al.\ 2011, \aap, 534, A83 


\bibitem[Ritchey et al.(2011)]{ritchey} Ritchey, A.~M., 
Federman, S.~R., \& Lambert, D.~L.\ 2011, \apj, 728, 36 



\bibitem[Scandariato et 
al.(2012)]{scandariato} Scandariato, G., Da Rio, N., Robberto, M., Pagano, I., \& Stassun, K.\ 2012, \aap, 545, A19 


\bibitem[Smith 
\& Brooks(2008)]{smithbrooks} Smith, N., \& Brooks, K.~J.\ 2008, Handbook of Star Forming Regions, Volume II, 5, 138 

{ \bibitem[Tan et al.(2006)]{tan06} Tan, J.~C., Krumholz, M.~R., \& McKee, C.~F.\ 2006, \apjl, 641, L121 }

  
\bibitem[Tognelli et 
al.(2011)]{tognelli} Tognelli, E., Prada Moroni, P.~G., \& Degl'Innocenti, S.\ 2011, \aap, 533, A109 




\bibitem[van Dokkum(2001)]{lacosmic} van Dokkum, P.~G.\ 2001, 
\pasp, 113, 1420 


\bibitem[Verma et 
al.(1994)]{verma94} Verma, R.~P., Bisht, R.~S., Ghosh, S.~K., et al.\ 1994, \aap, 284, 936 

\bibitem[Whittet(2003)]{whittet} Whittet, D.~C.~B.\ 2003, Dust in the galactic environment, 2nd ed.~ by D.C.B.~Whittet.~Bristol: Institute of Physics (IOP) Publishing, 2003 Series in Astronomy and Astrophysics, ISBN 0750306246.,  



\bibitem[Yonekura et al.(2005)]{yonekura05} Yonekura, Y., Asayama, 
S., Kimura, K., et al.\ 2005, \apj, 634, 476 

\end{thebibliography}
\end{document}